\documentclass{jfm}
\usepackage{graphicx}

\usepackage{upmath} 
\usepackage{amssymb} 

\usepackage{siunitx} 

\shorttitle{Taylor-Couette turbulence at radius ratio $\eta=0.5$}
\shortauthor{R.C.A. van der Veen and others}

\title{Taylor-Couette turbulence at radius ratio $\eta=0.5$: scaling, flow structures and plumes}

\author{
  Roeland C. A. van der Veen\aff{1},
  Sander G. Huisman\aff{1},
  Sebastian Merbold\aff{2},
  Uwe Harlander\aff{2},
  Christoph Egbers\aff{2},
  Detlef Lohse\aff{1}
 \and Chao Sun\aff{1,3}
   \corresp{\email{c.sun@utwente.nl}}
 }

\affiliation{
\aff{1}Physics of Fluids Group, MESA$^{+}$ Institute  and J.M. Burgers Centre for Fluid Dynamics, University of Twente, P.O. Box 217, 7500AE Enschede, The Netherlands
\aff{2}Department of Aerodynamics and Fluid Mechanics, Brandenburg University of Technology Cottbus-Senftenberg, Siemens-Halske-Ring 14, 03046 Cottbus, Germany
\aff{3}Center for Combustion Energy and Department of Thermal Engineering, Tsinghua University, Beijing 100084, China
}

\begin{document}

\maketitle

\begin{abstract}

Using high-resolution particle image velocimetry we measure velocity profiles, the wind Reynolds number and characteristics of turbulent plumes in Taylor-Couette flow for a radius ratio of 0.5 and Taylor number of up to $6.2\cdot10^9$. The extracted angular velocity profiles follow a log-law more closely than the azimuthal velocity profiles due to the strong curvature of this $\eta=0.5$ setup. The scaling of the wind Reynolds number with the Taylor number agrees with the theoretically predicted 3/7-scaling for the classical turbulent regime, which is much more pronounced than for the well-explored $\eta=0.71$ case, for which the ultimate regime sets in at much lower Ta. By measuring at varying axial positions, roll structures are found for counter-rotation while no clear coherent structures are seen for pure inner cylinder rotation. In addition, turbulent plumes coming from the inner and outer cylinder are investigated. For pure inner cylinder rotation, the plumes in the radial velocity move away from the inner cylinder, while the plumes in the azimuthal velocity mainly move away from the outer cylinder. For counter-rotation, the mean radial flow in the roll structures strongly affects the direction and intensity of the turbulent plumes. Furthermore, it is experimentally confirmed that in regions where plumes are emitted, boundary layer profiles with a logarithmic signature are created.

\end{abstract}


\section{Introduction}
\begin{figure}
\centerline{\includegraphics{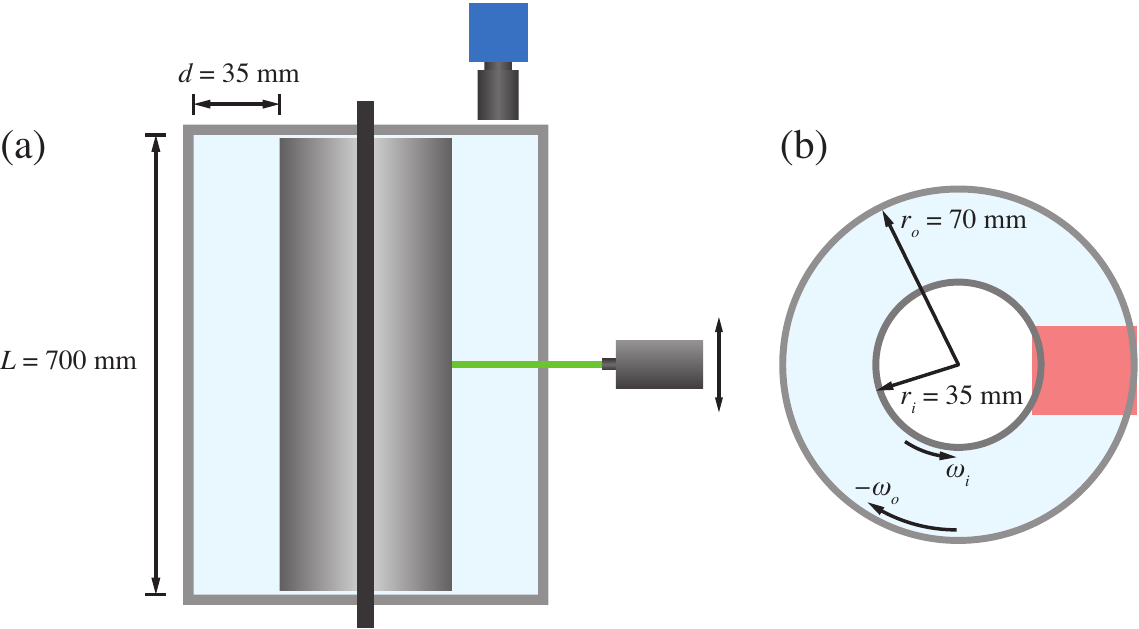}}%
 \caption{(a) Schematic of the vertical cross-section of the Taylor-Couette setup, showing the horizontal laser sheet which is imaged from above trough the transparent top plate. The laser can be traversed vertically. (b) Horizontal cross-section of the setup. The red rectangle represents the typical field of view.
 \label{fig:setup}}
 \end{figure}

The paradigmatic Taylor-Couette (TC) flow has long been a flow configuration of great interest to fluid dynamicists. It consists of flow between two coaxial cylinders that can independently rotate, see figure~\ref{fig:setup}. This system has been used extensively as a model in fluid physics because it is a closed system, has a relatively simple geometry and therefore has multiple symmetries.  After early investigations \citep{cou1890,ma1896, tay23}, \citet{wen33} started studying turbulence in this system. For a further historical overview, the reader is referred to \citet{don91}. For a review on TC flow at the onset of instabilities and slightly above, see \citet{far14}. The state-of-the-art of high Reynolds number Taylor-Couette turbulence is treated by \citefullauthor{gro16} (\citeyear{gro16}).

The two geometrical control parameters of TC flow are the ratio $\eta$ of the inner and outer cylinder radius and the aspect ratio $\Gamma=d/L$, where $d$ is the gap width and $L$ the height of the cylinders. In this work we use a small radius ratio of $\eta=0.5$, corresponding to a relatively wide gap. The vast majority of existing work (see overview by \citet{dub05} and review by \citet{gro16}) focuses on a radius ratio of 0.71 or higher. Lower $\eta$ experiments concern the Rayleigh-stable regime \citep{ji06}, mean flow and turbulence characteristics \citep{hou11} and global torque measurements \citep{mer13}. In addition, numerical work has been done for $\eta = 0.5$ \citep{don07,cho14,ost14pd}. The radius ratio is a key control parameter \citep{eck07b} in TC flow and it has been found that a low radius ratio Taylor-Couette system behaves differently as compared to higher $\eta$ setups in several aspects, which will be outlined below. The underlying reason for the different behaviour is the strong boundary layer asymmetry. Because the ratio of inner and outer boundary layer thicknesses scales as $\eta^3$ \citep{eck07b}, a strong asymmetry between the inner and outer boundary layer exists for small radius ratio.

As dimensionless control parameters of the system we use the (negative) rotation ratio $a=-\omega_o/\omega_i$ and the Taylor number \citep{eck07b}:
\begin{equation}
\text{Ta}=\frac{(1+\eta)^4}{64\eta^2} \frac{(r_o - r_i)^2 (r_i + r_o)^2 (\omega_i - \omega_o)^2}{\nu^2}
\end{equation}
where $\omega_{i,o}$  are the angular velocities of the inner and outer cylinder, $r_{i,o}$ are the inner and outer cylinder radii, $\eta=r_i/r_o$ is the radius ratio, and $\nu$ is the kinematic viscosity. The driving of the system can also be described by the shear Reynolds number \citep{dub05}:
\begin{equation}
\mathrm{Re_S}=\frac{2r_i r_o (r_o-r_i)}{\nu(r_o+r_i)}|\omega_o-\omega_i|,
\end{equation}
for which $\mathrm{Re_S}\propto\mathrm{Ta}^{1/2}$.

When increasing the driving strength (i.e. the Taylor number), TC flow first gradually undergoes a transition from a purely azimuthal, laminar state to a state where the bulk becomes turbulent while the boundary layers still remain laminar. The latter state is called the classical regime \citep{gro00}. By further increasing Ta, the ultimate regime \citep{kra62,gro11,he12} is reached, in which also the boundary layers are turbulent. The signature of these turbulent boundary layers are logarithmic velocity profiles which have recently been found for a radius ratio of $\eta = 0.716$ \citep{hui13} and for $\eta = 0.909$ \citep{ost15}. \citet{gro14} have derived theoretical velocity profiles, and found that the angular velocity profile follows a log-law more closely than the azimuthal velocity profile, an effect that is more pronounced with the stronger asymmetry for smaller $\eta$. In section \ref{subsec:azang} we set out to investigate the correspondence of experimentally measured angular and azimuthal velocity profiles to a log-law for $\eta=0.5$ and Taylor numbers up to $6.2\cdot10^9$ at the onset of the ultimate turbulent regime. Furthermore, the velocity profiles are compared to existing experimental and numerical work.

In addition to affecting the velocity profiles, the radius ratio strongly influences the transitional Taylor number for the ultimate regime \citep{rav10,mer13,ost14eta,ost14pd}. For $\eta = 0.5$ the ultimate regime does not start before $\mathrm{Ta} = 10^{10}$ \citep{mer13,ost14pd}, whereas the transition for a higher radius ratio of $\eta=0.71$ already occurs at $\mathrm{Ta} = 5\cdot10^8$ \citep{gil12,ost14pd}. Very different scaling of the angular momentum transfer and the ``wind'' in the gap with the driving paramater Ta exist for the classical and ultimate regime \citep{gro00,gro11}. The late onset of the ultimate regime for $\eta=0.5$ gives us the opportunity to analyze the scaling of the wind Reynolds number with Taylor numbers in the classical TC regime, up to the onset of the ultimate regime and compare it with theoretical predictions (section \ref{subsec:rewind}).

Another area where the influence of the radius ratio is apparent, is the angular momentum transport trough the gap. This transport is a key response parameter of the system and is directly connected to the torque required to maintain constant cylinder velocities \citep{eck07b}. The radius ratio strongly influences the rotation ratio for which optimal momentum transport occurs \citep{gil12,mer13,ost14eta}. For example, the optimal momentum transport occurs at $a_{\mathrm{opt}}\approx0.33$ for $\eta=0.714$ \citep{gil12} whereas it is $a_{\mathrm{opt}}\approx0.20$ for $\eta=0.5$ \citep{mer13}.

Roll structures play an important role in the momentum transport in the Taylor-Couette system \citep{fen79,and86,lew99,hui14,ost14bl}. With increasing Taylor number, the system undergoes transitions from a purely laminar state to having steady, coherent Taylor vortices, to having modulated, wavy Taylor vortices which eventually evolve into chaotic turbulent Taylor vortices \citep{lew99}. But even in the ultimate regime, for some $a>0$, it was found that remnants of these rolls are present in time-averaged quantities \citep{tok11,hui14,ost14bl}, whereas for $a=0$ (pure inner cylinder rotation) these rolls vanish for large Taylor number \citep{lat92}. In section \ref{subsec:rolls} we will characterise the roll structures for $\eta=0.5$ at both pure inner cylinder rotation $a=0$ and slight counter-rotation $a=0.2$, at which optimal transport occurs.

It was recently proposed that the logarithmic velocity profiles in ultimate TC flow are triggered by turbulent plume ejection \citep{ahl14,ost14bl,poe15}, which in turn is regulated by the relative strength of the axial and radial mean flow. Using time-resolved azimuthal and radial velocity field measurements at several heights in the setup we will investigate the connection between the roll structures that generate strong radial and axial flow, and the turbulent plumes that emanate from either the inner or outer cylinder (section \ref{subsec:plumes}). Futhermore, in section \ref{subsec:logprofiles} we will quantify how these plumes affect the logarithmic nature of the velocity profiles.

  \begin{figure}
 \centerline{\includegraphics[scale=0.5]{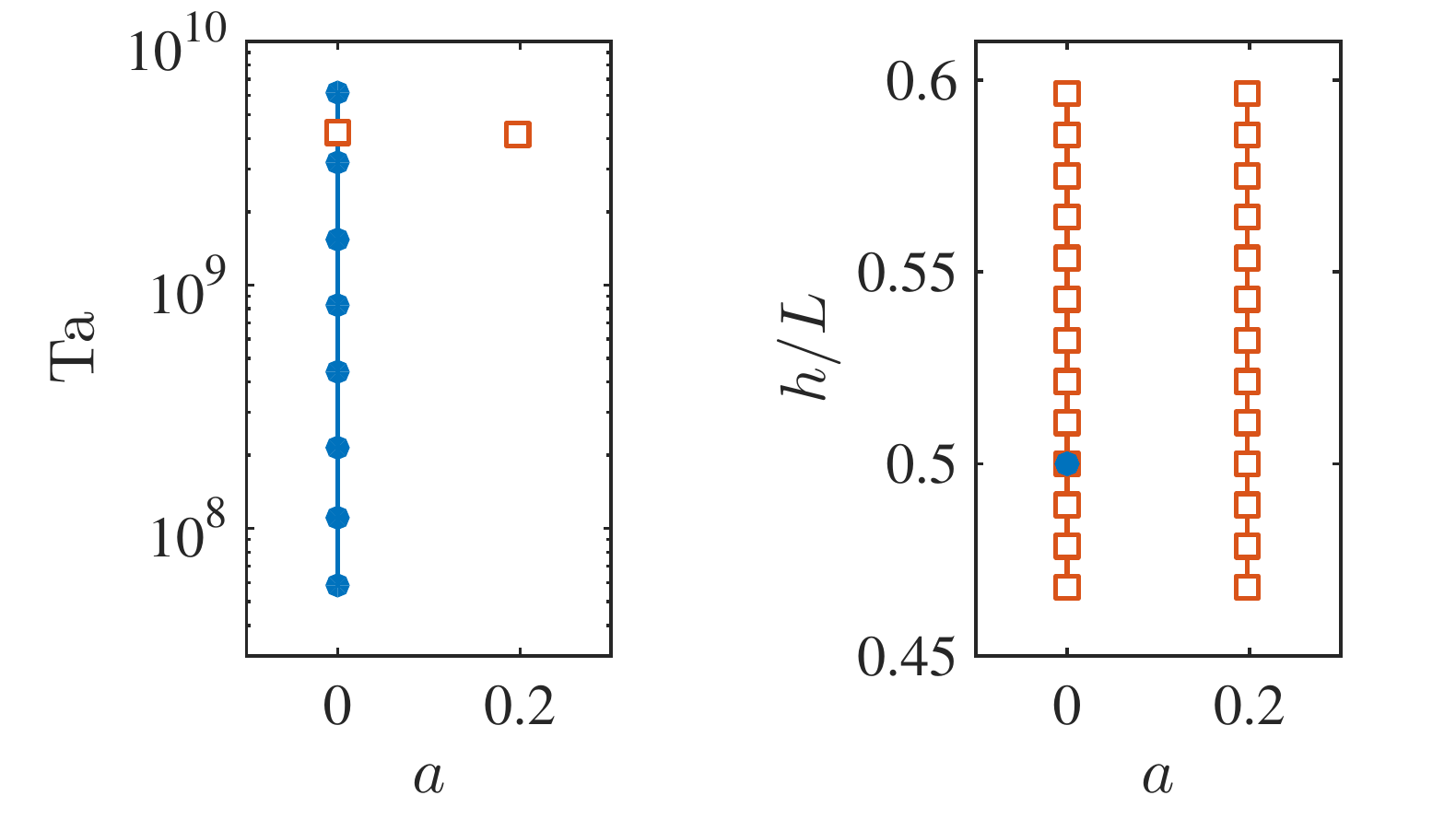}}%
 \caption{The explored parameter space. The Taylor number ($\mathrm{Ta}$) is varied for inner cylinder rotation ($a=0$), and axial scans have been performed for  (negative) rotation ratios $a=0$ and $a=0.2$.
 \label{fig:params}}
 \end{figure}

\section{Setup \& explored parameter space}
\subsection{Setup}

The experiments were carried out in the Cottbus Taylor-Couette facility \citep{mer11,mer13}. The inner and outer cylinder radii are $r_i = 35.0\pm0.2 \mathrm{mm}$ and $r_o = 75.0\pm0.2 \mathrm{ mm}$, respectively, the height of the setup is $L = \SI{700}{ mm}$. This gives a radius ratio of $\eta = 0.5$ and an aspect ratio of $\Gamma = 20$. The cylindricities of the cylinders that were used are $\SI{0.4}{ mm}$ and $\SI{0.3}{ mm}$ for the inner and outer cylinder, respectively. The maximum rotation rates are 5 Hz for both the inner and outer cylinder.

The end plates rotate with the outer cylinder. The outer cylinder and the top plate are transparent, making the setup ideally suited to be used in combination with particle image velocimetry (PIV). A high-resolution PIV camera (LaVision Imager sCMOS) with a resolution of 2560 $\times$ 2160 pixels and a frame rate of 50 Hz is installed above the top end plate, pointing downwards. Water is used at the working fluid ($20^{\circ}\mathrm{C}$, $\nu=\SI{1.0e-6}{m^2/s}$). The water contains fluorescent particles (Dantec Dynamics, PMMA-RhB, 1-20 $\mu\mathrm{m}$) with a maximum Stokes number of $\mathrm{St} = \tau_p/\tau_\eta \approx 10^{-4}\ll1$, which means that they faithfully follow the flow and can be considered as tracer particles. The flow is illuminated by a horizontal light sheet from a high-powered pulsed Nd:YLF dual cavity laser (Litron LDY303HE). Because of the high-resolution PIV camera, very high resolution measurements of the flow fields can be achieved. The imaging of the full width of the gap combined with a vector grid of 16 $\times$ 16 pixels with 50\% overlap results in a velocity vector spacing of 0.13 mm. The PIV system is operated in dual frame mode, allowing for an interframe time $\Delta t$ smaller than the inverse frame rate. The PIV image pairs are processed using LaVision DaVis software, after which the flow fields are transformed to the radial velocity $u_r(\theta,r,z,t)$ and the azimuthal velocity $u_\theta(\theta,r,z,t)$.

 \subsection{Explored parameter space}

The parameter space that was explored for this work is shown in figure~\ref{fig:params}. The first set of measurements that is treated here consists of varying the Taylor number between $\mathrm{Ta} = 5.8\cdot10^7$ and $\mathrm{Ta} = 6.2\cdot10^9$ (corresponding to a shear Reynolds number of $\mathrm{Re_S} =  6.0\cdot10^3$ to $6.2\cdot10^4$) for inner cylinder rotation only ($a=0$) at mid-height ($h = 0.5L$). These measurements were performed as a single continuous experiment; the cylinder velocity was increased slowly between experiments and before each measurement approximately 10 minutes were taken to give the flow ample time to stabilise. Each measurement in this set consists of 10000 PIV image pairs, which were recorded at 25 frames per second. This corresponds to 235 and 2418 cylinder revolutions for $\mathrm{Ta} = 5.8\cdot10^7$ and $\mathrm{Ta} = 6.2\cdot10^9$, respectively.

In addition to investigating the Taylor number dependence, the height of the laser sheet was varied to characterise the axial dependence of the flow field. At 13 different heights with 7.5 mm spacing, 5000 image pairs of the velocity field were recorded. This was done at $\mathrm{Ta} = 4.2\cdot10^9$ for both pure inner cylinder rotation and optimal \citep{mer13} counter-rotation $a=0.2$. 

  \begin{figure}
 \centerline{\includegraphics[width=384pt]{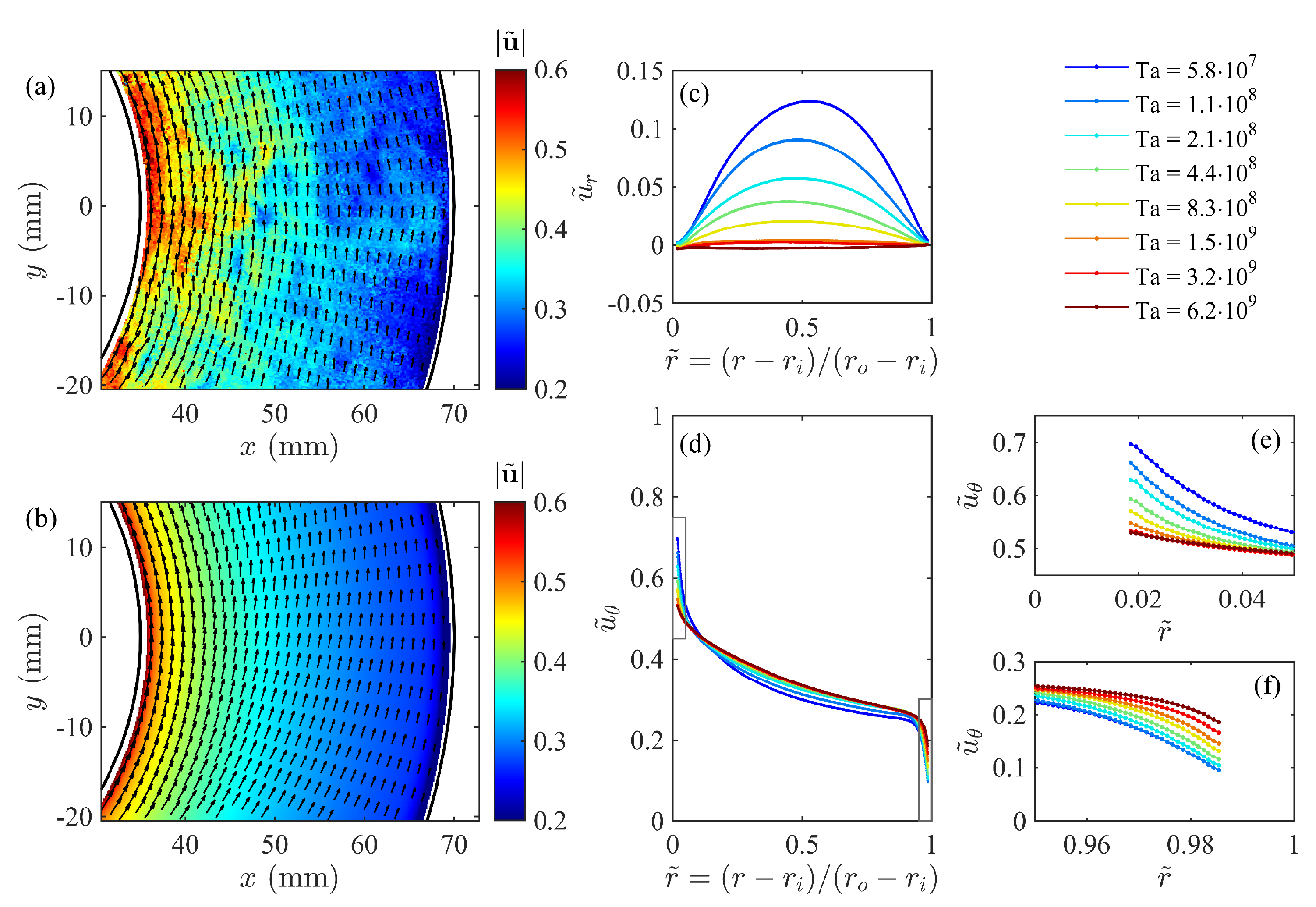}}
 \caption{Overview of the flow profiles for varying Taylor number with pure inner cylinder rotation $a = 0$ at mid-height $h/L=0.5$.
 (a) Snapshot of the flow field for $\mathrm{Ta} = 6.2\cdot10^9$. The colours and lengths of the arrows indicate the norm of the velocity $\vert\tilde{\mathbf{u}}\vert$.
 (b) Averaged flow field over 10000 PIV image pairs at $\mathrm{Ta} = 6.2\cdot10^9$.
 (c) Radial velocity profiles across the gap of the TC apparatus, normalised by the inner cylinder velocity. For lower Taylor numbers there is still a strong radial flow, which can be attributed to the presence of Taylor vortices.
 (d) Azimuthal velocity profiles normalised by the inner cylinder velocity. For increasing Taylor number, the profiles become flatter and the boundary layers steeper.
 (e) and (f) Magnification of the azimuthal velocity profiles in (d) close to the cylinders.
 \label{fig:profiles}}
 \end{figure}  

  \begin{figure}
 \centerline{\includegraphics[width=384pt]{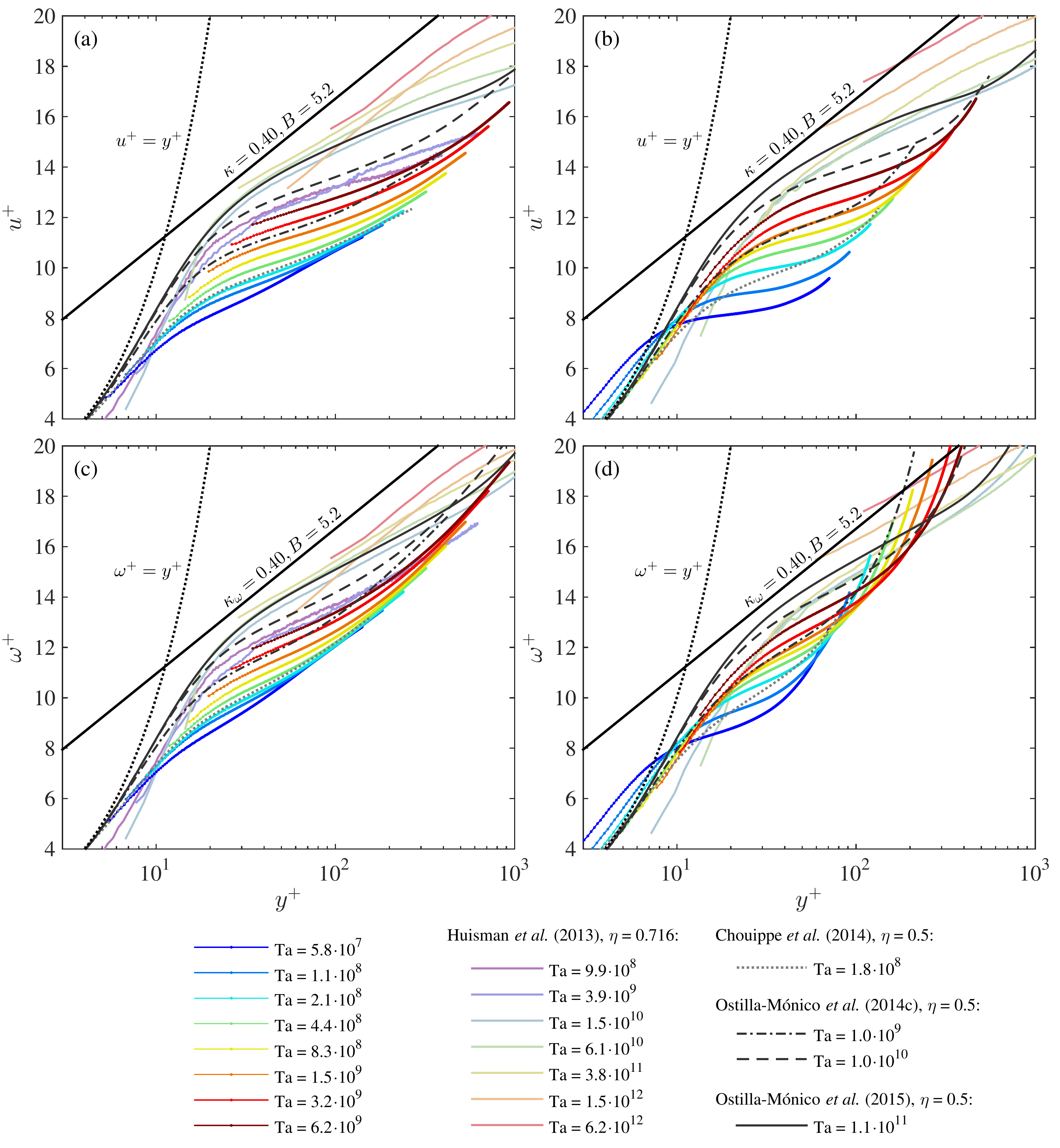}}%
 \caption{ Velocity profiles for inner cylinder rotation $a=0$ and varying Taylor number.
 (a) Azimuthal velocity profiles near the inner cylinder ($\tilde{r} \in [0,1/2]$). The velocity $u^+$ is defined as $(u_\theta(r_i)-u_\theta(r))/u_{\tau,i}$ with the friction velocity $u_{\tau,i}=\sqrt{\tau_{w,i}/\rho}$ containing the wall shear stress $\tau_{w,i}=T/2\pi r_i^2L$ with the torque $T$.
 The wall normal distance is scaled using the viscous length scale $\delta_{\nu,i}=\nu/u_{\tau,i}$. 
 (b) Azimuthal velocity profiles near the outer cylinder ($\tilde{r} \in [1/2,1]$). For the outer cylinder case, the velocity $u^+$ is defined as $(u_\theta(r)-u_\theta(r_o))/u_{\tau,o}$ with $u_{\tau,o}=\sqrt{\tau_{w,o}/\rho}$ and $\tau_{w,o}=T/2\pi r_o^2L$. The normalised wall normal distance is $y^+=(r_o-r)/\delta_{\nu,o}$ with $\delta_{\nu,o}=\nu/u_{\tau,o}$.
(c) Angular velocity profiles near the inner cylinder ($\tilde{r} \in [0,1/2]$). The velocity $\omega^+$ is defined as $(\omega(r_i)-\omega(r))/(u_{\tau,i}/r_i)$
(d) Angular velocity profiles near the outer cylinder ($\tilde{r} \in [1/2,1]$). For the outer cylinder case, the velocity $\omega^+$ is defined as $(\omega(r)-\omega(r_o))/(u_{\tau,o}/r_o)$
All figures include the logarithmic law of the wall from Prandtl and von K\'{a}rm\'{a}n $u^+=1/\kappa \ln{y^+} +B$  with the typical values of $\kappa=0.40$ and $B=5.2$ (see \citet{mar13} and references therein), the viscous sublayer $u^+ = y^+$, DNS data from \citet{cho14}, \citet{ost14pd} and \citet{ost15} at $\eta=0.5$ and measurement data from \citet{hui13} at $\eta=0.716$.
 \label{fig:profiles2}}
 \end{figure}

\section{Results}
\subsection{Azimuthal and angular velocity profiles}
\label{subsec:azang}

In this part the properties of flow profiles for varying Taylor number will be investigated. An example of an instantaneous flow field is shown in figure~\ref{fig:profiles}(a), while the averaged field is shown in (b). From (c) it is apparent that at this specific axial position (mid-height) a significant mean radial flow still exists for the lower Taylor numbers up to approximately $10^9$. These are Taylor vortices, which at $a=0$ disappear for higher Taylor numbers \citep{lat92}. With increasing Taylor number the azimuthal velocity profile in the bulk becomes flatter (figure~\ref{fig:profiles}(d-f)), but it is still apparent that the strong curvature of a radius ratio of $\eta=0.5$ creates a significant asymmetry between the inner and outer boundary layer.

In figure~\ref{fig:profiles2} profiles up to the middle of the gap are shown, normalised to the wall normal distance $y^+$ from both the inner and outer cylinder, and to the azimuthal velocity $u^+$ ((a) and (b)) and angular velocity $\omega^+$ ((c) and (d)). The friction velocity $u_{\tau}$ that is used in this normalisation is calculated with global torque data from \citet{mer13}.

As the driving is increased, the profiles slowly approach the Prandtl-von K\'{a}rm\'{a}n log-law, although at these Taylor numbers the log layer is not yet fully developed. For comparison, profiles at higher Taylor numbers and aspect ratio of $\eta=0.716$ from \citet{hui13} are also plotted. The data show good agreement with direct numerical simulations (DNS) at the same radius ratio $\eta=0.5$ from \citet{cho14} at $\mathrm{Ta}=1.8\cdot10^8$, \citet{ost14pd} at $\mathrm{Ta}=1.0\cdot10^9$ and $\mathrm{Ta}=1.0\cdot10^{10}$ and \citet{ost15} at $\mathrm{Ta}=1.1\cdot10^{11}$.

When comparing the azimuthal velocity $u^+$ to the angular velocity $\omega^+$ (i.e. (a) to (c) and (b) to (d)) it can be seen that the angular velocity profiles curve upwards more compared to the azimuthal ones. This is consistent with the theoretical argument of \citet{gro14} that the angular velocity profile $\omega^+$ is closer to a log-law than the azimuthal velocity profile $u^+$.  The effect is much more pronounced for the smaller $\eta=0.5$ here as compared to the $\eta=0.716$ from \citet{hui13}. Note that, for example, the lower Taylor number $u^+$ profiles at $\eta=0.716$ are above the profiles measured here, but that they cross when represented as $\omega^+$. This clearly shows the influence of the large curvature for small radius ratio TC setups.


 \begin{figure}
 \centerline{\includegraphics[scale=0.4444]{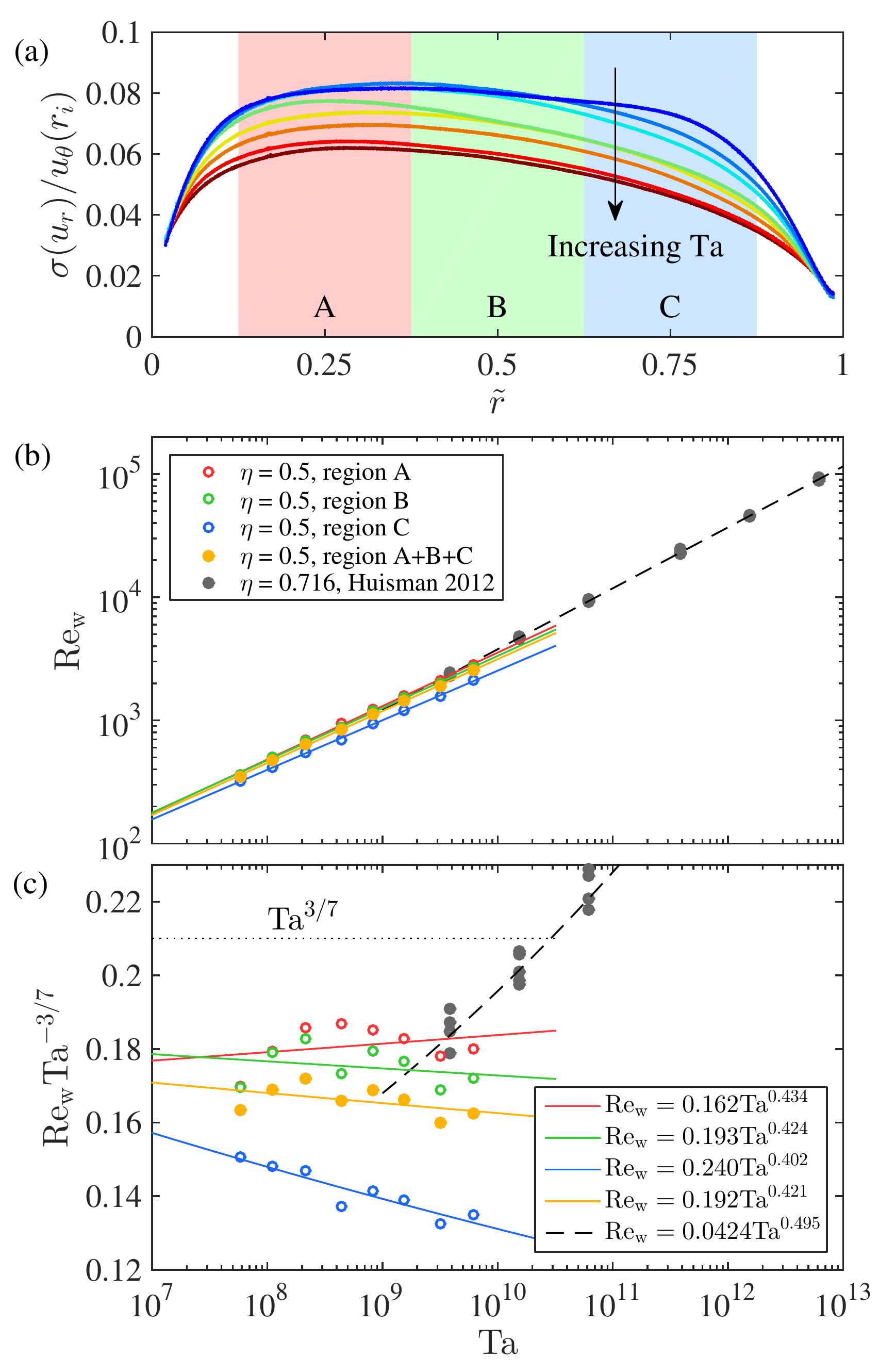}}
 \caption{ The scaling of the wind Reynolds number with Taylor number for inner cylinder rotation $a=0$.
 (a) The standard deviation of the radial velocity over 10000 frames (400 s) and the azimuthal direction, normalised by the inner cylinder velocity, as a function of the normalised radial position. See the legend in figure~\ref{fig:profiles} or \ref{fig:profiles2} for the values of the Taylor numbers. The three areas A ($\tilde{r} \in [1/8,3/8]$) , B ($\tilde{r} \in [3/8,5/8]$) and C ($\tilde{r} \in [5/8,7/8]$) indicate regions over which $\sigma$ is averaged.
 (b) $\mathrm{Re_w}$ versus $\mathrm{Ta}$ averaged over $\tilde{r}$-values corresponding to regions A, B, C and all three combined ($\tilde{r} \in [1/8,7/8]$). In addition, data from \citet{hui12} with their fit are shown. This data uses the equivalent to area B for averaging.
  (c) Data in (b) compensated by $\mathrm{Ta}^{3/7}$. The fits are $\mathrm{Re_w} = 0.162\mathrm{Ta}^{0.434}$ for area A, $\mathrm{Re_w} = 0.193\mathrm{Ta}^{0.424}$ for area B, $\mathrm{Re_w} = 0.240\mathrm{Ta}^{0.402}$ for area C and $\mathrm{Re_w} = 0.192\mathrm{Ta}^{0.421}$ for the three areas combined.
 \label{fig:Rew}}
 \end{figure}

\subsection{Wind Reynolds number}
\label{subsec:rewind}

The degree of turbulence of the wind in the gap of the cylinders, which measures the strength of the secondary flows $u_r$ and $u_z$, can be characterised by the wind Reynolds number. We use the standard deviation of the radial velocity to quantify the wind Reynolds number:
\begin{equation}
\mathrm{Re_w} = \sigma(u_r)d/\nu
\end{equation}
with the gap width $d = r_o-r_i$. In the analogy between TC and Rayleigh-B\'{e}nard flow, the unifying theory of \citet{gro00} predicts that the wind Reynolds number in the classical turbulent regime scales as:
\begin{equation}
\mathrm{Re_w} \propto \mathrm{Ta}^{3/7}.
\end{equation}
In contrast, in the ultimate turbulent regime the scaling was found to be $\mathrm{Re_w} \propto \mathrm{Ta}^{1/2}$ \citep{gro11}, which was confirmed experimentally by \citet{hui12} for TC flow with $\eta=0.716$ for $\mathrm{Ta}$ up to $6.2\cdot10^{12}$.

The highest Taylor number we achieve in the present measurements is $\mathrm{Ta}=6.2\cdot10^9$, which for $\eta=0.5$ with the higher transitional Taylor number of $\mathrm{Ta} = 10^{10}$ \citep{mer13,ost14pd} implies that we are in the classical turbulent regime. Therefore we expect the theoretically predicted 3/7-scaling to hold. This classical scaling was measured in Rayleigh-B\'{e}nard convection \citep{he12}, but to our knowledge has not yet been confirmed for TC flow.

The wind Reynolds number $\mathrm{Re_w}$ is extracted from the velocity field $u_r(\theta,r,z,t)$ by calculating the standard deviation $\sigma(u_r(r,\theta,t))$ over time and the azimuthal direction, which is then averaged over a certain range in the gap, see figure~\ref{fig:Rew}. When looking at the shape of the $\tilde{r}$-dependence of $\sigma$, it is apparent that an asymmetry is present. The fluctuations exhibit a maximum around a quarter of the gap width away from the inner cylinder. It is likely that the strong curvature of the $\eta=0.5$ setup causes this assymmetry. Calculating the wind Reynolds number thus comes with an arbitrariness in the area to choose for averaging the profile of $\sigma(r)$, although excluding the boundary layers is reasonable because the interest lies in the convective bulk transport, as opposed to the diffusive transport in the boundary layers. The dependence of $\mathrm{Re_w}(\mathrm{Ta})$ is shown in figure~\ref{fig:Rew}(b) for different regions of averaging.

From figure~\ref{fig:Rew}(c) it can be seen that a clear power law scaling is present with an exponent that ranges from 0.402 to 0.434, depending on the region of averaging. This is consistent with the theoretically predicted classical regime scaling of $3/7\approx0.429$ \citep{gro00}.

 \begin{figure}
 \centerline{\includegraphics[width=384pt]{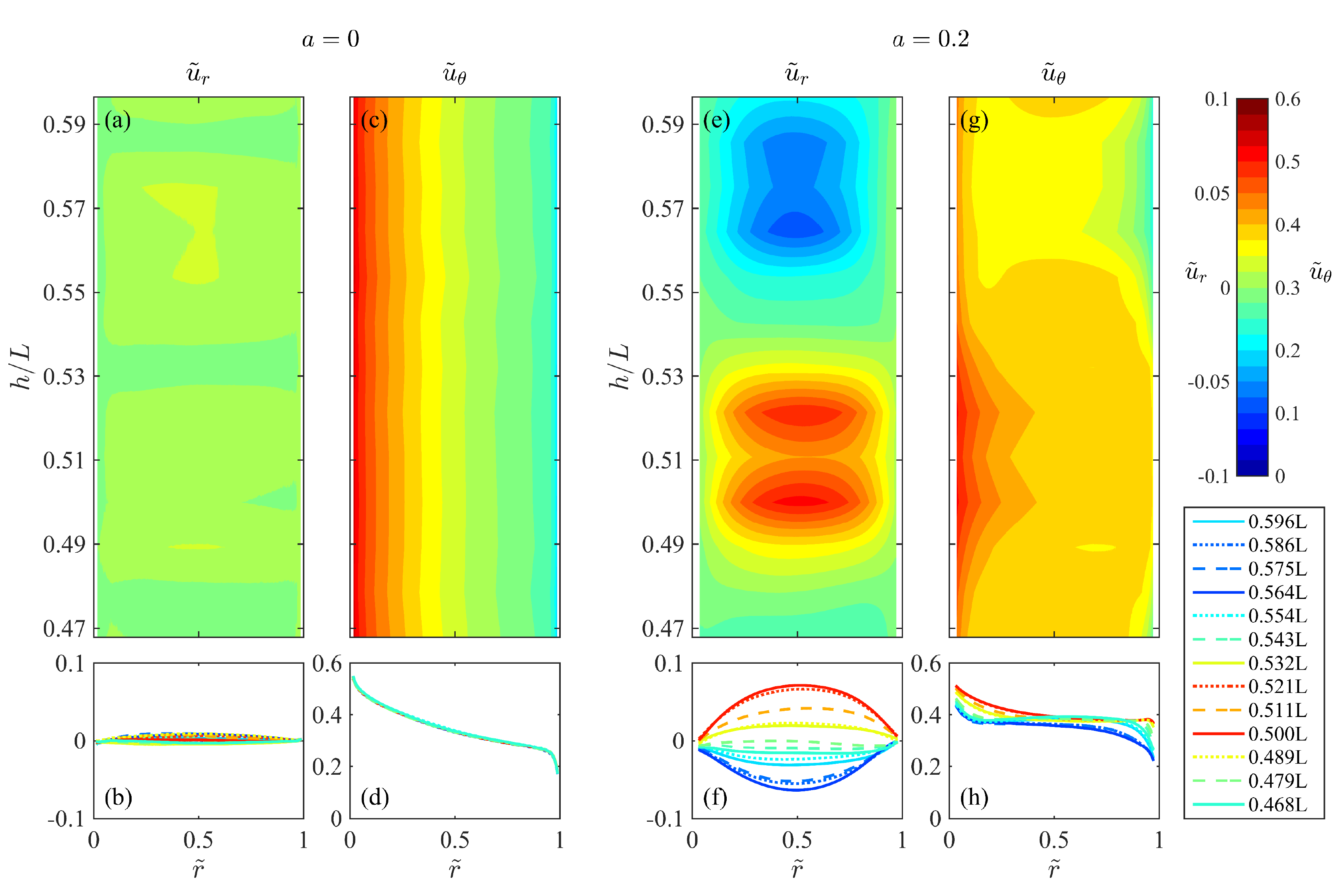}}
 \caption{
 Dependence of flow profiles on axial position for rotation ratios $a=0$ (a-d) and $a=0.2$ (e-h) for $\mathrm{Ta}=4.2\cdot10^9$. The normalised radial velocity is $\tilde{u}_r=u_r(r)/(u_i-u_o)$ and the azimuthal velocity is $\tilde{u}_{\theta}=(u_{\theta}(r)-u_o)/(u_i-u_o)$. The data is represented in a colour map (a,c,e,g) and as profiles in (b,d,f,h). For the visualisation of the colour map, bilinear interpolation is used.
 \label{fig:height}}
 \end{figure}

\subsection{Roll structures}
\label{subsec:rolls}

The phenomena of roll structures, turbulent plumes and logarithmic velocity profiles in Taylor-Couette flow are intimately connected \citep{ost14bl,ost14pd}. We will first show an example of roll structures that are observed in TC flow. Flow profiles at several heights for both $a = 0$ and $a = 0.2$ are measured, allowing us to visualise roll structures as shown in figure~\ref{fig:height}. At this reasonably high Taylor number of $\mathrm{Ta}=4.2\cdot10^9$, no clear roll structures exist for inner cylinder rotation only. The radial velocity shows some very faint indications of rolls, with a maximum value of $\tilde{u}_r=0.01$ in the center of the gap. In contrast, in the case of counter-rotation ($a = 0.2$), very strong roll structures can be seen. The fact that there are strong rolls at optimal counter-rotation and no clear structures at inner cylinder rotation only, is corroborated by recent work for higher radius ratio \citep{tok11,hui14,ost14bl}.

In the radial flow, between two rolls at approximately $h/L=0.51$, there is very strong outwards radial flow (see figure \ref{fig:height}(e)). Specifically, the height 0.52L corresponds to the bottom of a roll, while 0.50L corresponds to the top of the roll below. Exactly in between the rolls at $0.51L$, the radial flow surprisingly exhibits a local minimum. In figure~\ref{fig:height}(g,h) it can be seen that in the case of positive (outward) radial flow, inner cylinder velocity is advected outwards, shifting the profiles upwards at the inner half of the gap. For negative (inward) radial flow, the profiles shift downwards in the outer half of the gap. In summary, the roll structures are represented by a large secondary flow of $u_r$, which advects velocity from the cylinders and changes the azimuthal velocity profiles. In the following two subsections we will elaborate on the characteristics of turbulent plumes and their effect on the velocity profiles.

 \subsection{Turbulent plumes}
\label{subsec:plumes}

  \begin{figure}
 \centerline{\includegraphics[width=384pt]{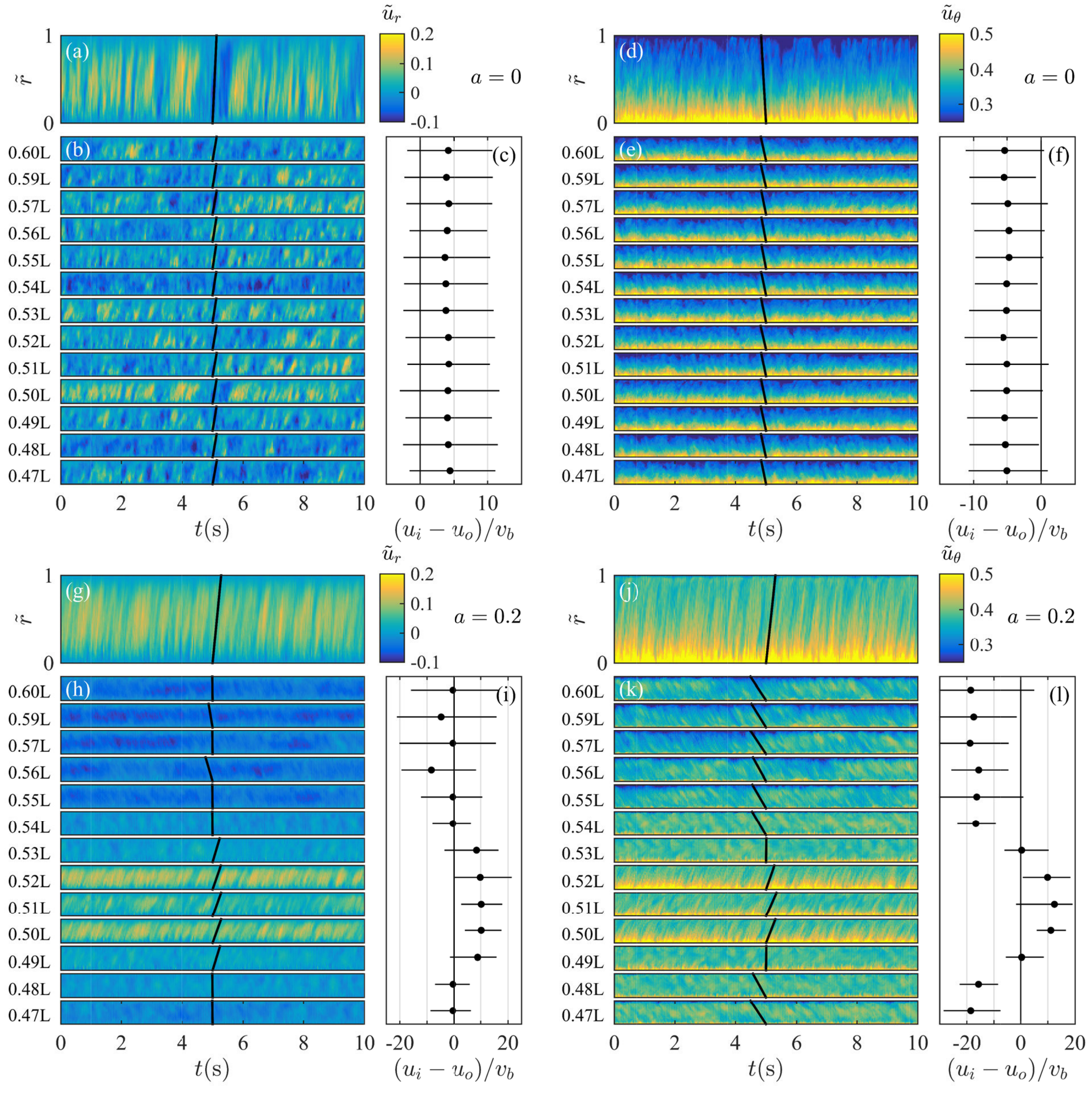}}%
 \caption{
 Overview of turbulent burst velocity depending on axial position, for both pure inner cylinder rotation $a=0$ (a-f) and counter-rotation of $a=0.2$ (g-l), at $\mathrm{Ta}=4.2\cdot10^8$. The radial velocity dependence is shown in (a,b,c,g,h,i) and the azimuthal velocity in (d,e,f,j,k,l).
(a,d,g,j) Example of the radial and azimuthal velocity as a function of the normalised gap distance and time for the axial position of $0.50L$. Only 10 s of the total 200 s per experiment are shown for clarity. Diagonal lines with a positive (negative) slope correspond to plumes coming from the inner (outer) cylinder. The extracted mean plume velocity (see text for method) is shown as a black line.
(b,e,h,k) As above, for every height normalised by the total height of the cylinder $L$.
(c,f,i,l) The inverse of the resulting normalised burst velocity. For $a=0$ the inner cylinder velocity is \SI{1.1}{m/s} and for $a=0.2$ the inner and outer cylinder velocities are \SI{0.91}{m/s} and \SI{-0.36}{m/s}, respectively. The error bars show the width of the test function at 0.95 times the peak value.
 \label{fig:bursts1}}
 \end{figure} 

In turbulent Taylor-Couette flow, structures detach from the boundary layers, called herring-bone streaks \citep{don07} or velocity plumes \citep{ost14bl}, in correspondence to the thermal plumes in the analogous turbulent Rayleigh-B\'{e}nard convection. These plumes in TC flow are large-scale spatial and temporal fluctuations in the velocity fields that detach from either the inner or outer cylinder and advect velocity from the respective cylinder. Because of the sufficiently high frame rate of 25 fps and large amount of data recorded in our experiments, these plumes can be resolved in time, and the typical velocity can be extracted. We are interested in understanding how these plumes are affected by roll structures and whether some scaling relation exists for the typical plume velocity as a function of the driving parameter $\mathrm{Ta}$.
 
The time dependence of the profiles of the azimuthal and radial velocity can be represented by the spatio-temporal fields $u_r(r,t)$ and $u_{\theta}(r,t)$, with the value of the velocities represented by a colour map, as shown in \textit{e.g.} figure~\ref{fig:bursts1}(a). In this way, structures or fluctuations of the profiles that propagate inwards or outwards become visible as diagonal lines. It has to be stressed that these diagonals do not represent a nonzero mean radial flow, but only the fluctuations on these profiles that travel outward or inward.
In order to assign a velocity to these plumes we employ an analysis which is based on applying the following affine shearing transformation to the functions $u_r(r,t)$ and $u_{\theta}(r,t)$:
\begin{equation}
(t,\tilde{r})\mapsto(t+d\cdot\tilde{r}/v_b,\tilde{r})
\end{equation}
with $v_b$ the plume velocity. For a certain amount of shearing, the plumes which were represented by diagonal lines, become vertical lines. To find this optimal value of shearing, we postulate that when averaging the sheared function over $\tilde{r}$, the standard deviation of the resulting time signal has a maximum as a function of the shearing value. At this optimal shearing, the plumes are represented by narrow peaks in the time signal, creating a high standard deviation. Using a golden section search algorithm \citep{ki53} the optimal shearing value is found, which corresponds to the mean velocity of the plumes $v_b$.

   \begin{figure}
 \centerline{\includegraphics[scale=0.5]{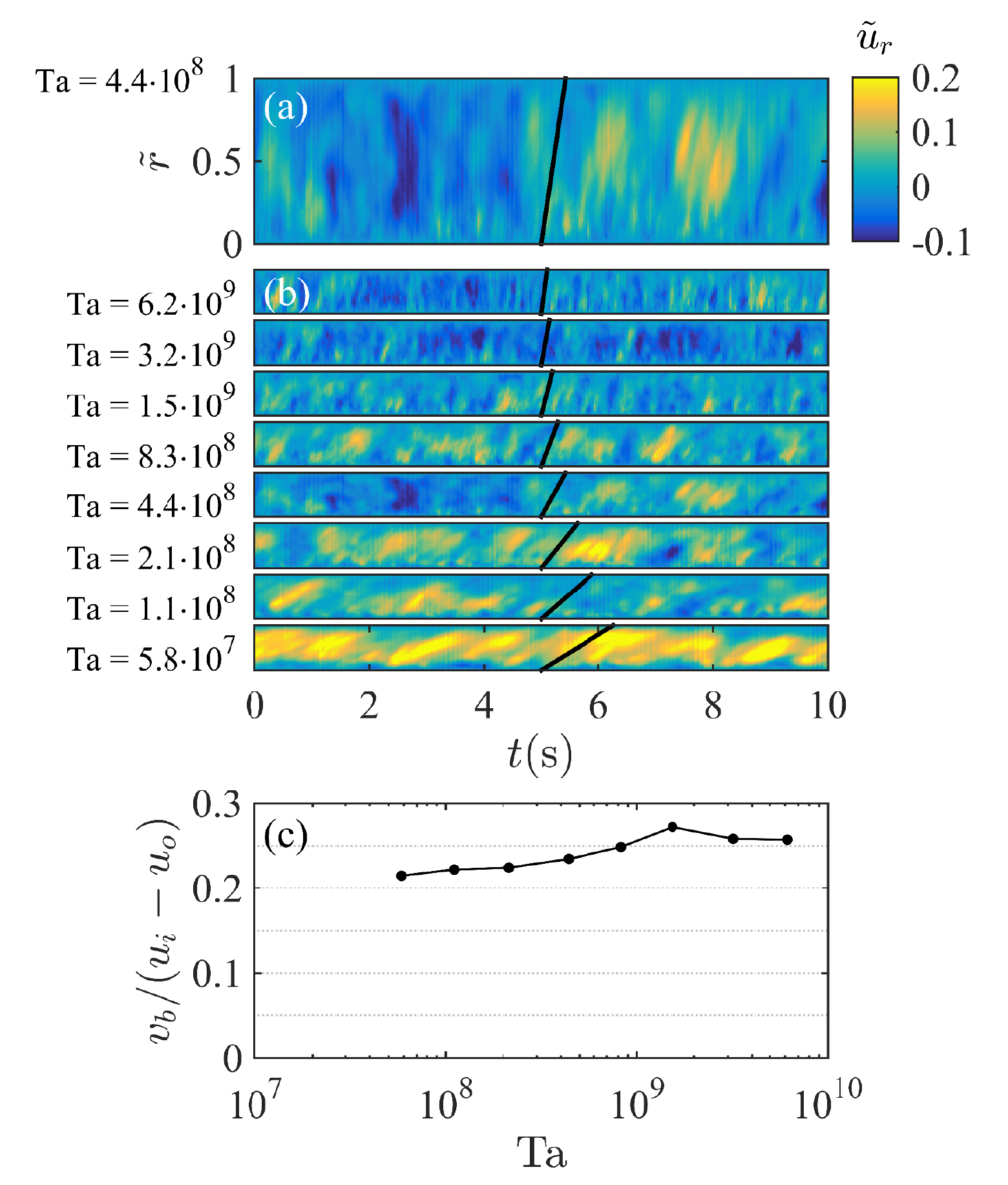}}
 \caption{Turbulent plumes in the radial velocity for varying  $\mathrm{Ta}$ and $a=0$. (a) and (b) as in figure~\ref{fig:bursts1}. (c) The plume velocity normalised by the inner cylinder velocity ($u_o=0$). Errors are of the same order as in figure~\ref{fig:bursts1}(c).
 \label{fig:bursts2}}
 \end{figure} 

The dependence of the plume velocity on the axial position and the rotation ratio is shown in figure~\ref{fig:bursts1}. Again, for inner cylinder rotation only ($a=0$), no height dependence can be seen. The fluctuations on the radial velocity are moving outwards with a velocity $v_b/(u_i-u_o)=0.25\pm0.02$ ($u_o = 0$, $2\sigma$ error). When the azimuthal velocity is considered, the picture looks different. Crossing diagonal lines can be seen, meaning that plumes emit from both the inner and outer cylinder. Despite of this, a clear peak was found for inward plumes with an average relative velocity of $v_b/(u_i-u_o)=0.20\pm0.02$. The velocity of these plumes is of the same order, but it remains to be explained why the main plume direction is different for the two velocity components.

For counter-rotation ($a=0.2$), the bottom half of figure~\ref{fig:bursts1} clearly indicates that roll structures are present. There is a correspondence between the sign of the mean radial velocity and the direction of these turbulent plumes, coming from either the inner or outer cylinder. Compare figure~\ref{fig:height}(e) with figure~\ref{fig:bursts1}(h,k). For both velocity components, the direction of the plumes corresponds to the direction of the mean radial velocity. The mean radial velocity forces the plumes strongly in the inward or outward direction, also typically creating a higher plume velocity than for the $a=0$ case. Whereas for the $a=0$ case there was some ambiguity in the direction of the plumes for the azimuthal velocity, for $a=0.2$ it exactly follows the roll structure. When looking at the radial velocity, the inward plumes are less pronounced than the outward ones, likely due to a tendency for the $u_r$ plumes to flow outwards, as could be seen for the $a=0$ case.
 
In addition to the dependence on the axial position and the rotation ratio, in figure~\ref{fig:bursts2} the dependence of the plume velocity on the driving parameter $\mathrm{Ta}$ is shown. With increasing driving strength, the velocity of the plumes also increases. When the plume velocity is normalised by $(u_i-u_o)$ it is however nearly constant over two decades of $\mathrm{Ta}$.
 
   \begin{figure}
\centerline{\includegraphics[width=384pt]{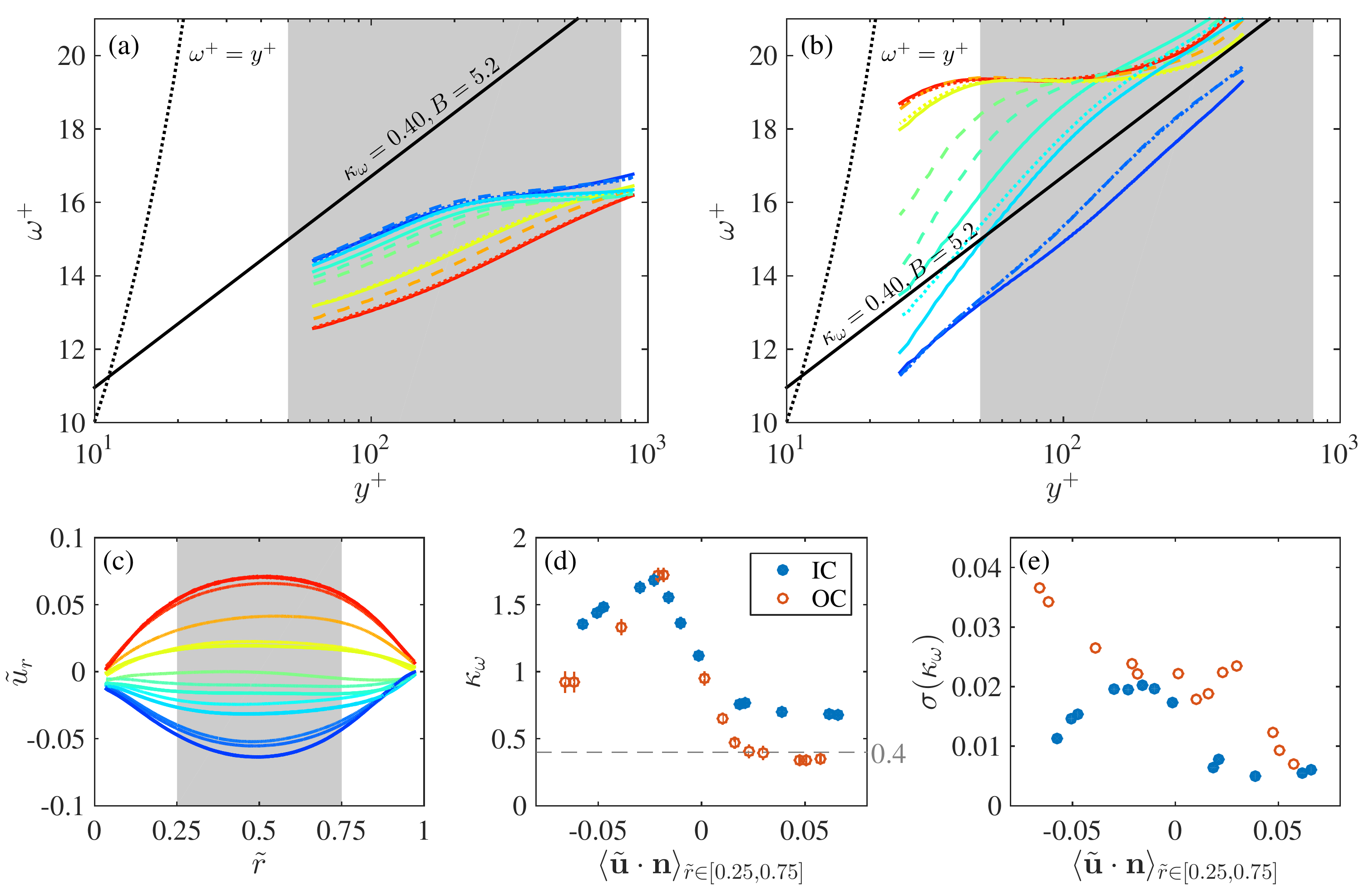}}%
 \caption{Angular velocity profile dependence on mean radial flow for counter-rotation $a=0.2$ and $\mathrm{Ta}=4.2\cdot10^9$ at varying axial positions (data from figure \ref{fig:height}(h)). (a) Angular velocity profiles near the inner cylinder ($\tilde{r} \in [0,1/2]$). The angular velocity $\omega^+$ and wall normal distance $y^+$ are defined in the caption of figure \ref{fig:profiles2}, the colours and line styles of the profiles correspond to those in figure \ref{fig:height}. The gray area is the fitting domain $y^+\in [50,800]$, chosen to encompass the start of the outer region up to mid-gap for both inner and outer cylinder. (b) Angular velocity profiles near the outer cylinder ($\tilde{r} \in [1/2,1]$). (c) Normalised radial velocity $\tilde{u}_r$ as in figure \ref{fig:height}(f). (d) The von K\'{a}rm\'{a}n constant $\kappa_{\omega}$ found by fitting $\omega^+=1/\kappa_{\omega} \ln{y^+} +B$ for $y^+\in [50,800]$ near the inner (IC) and outer cylinder (OC). The horizontal axis shows the normalised wall normal velocity, equal to $\tilde{u}_r$ and $-\tilde{u}_r$ for the IC and OC respectively, averaged over $\tilde{r}\in [1/4,3/4]$. The error bars have a total length of four times the standard error of the fit. (e) The standard error of the fitting parameter $\kappa_{\omega}$. 
 \label{fig:kappa}}
 \end{figure}

 \subsection{Logarithmic velocity profiles} 
\label{subsec:logprofiles} 
 
It has been proposed that the logarithmic temperature profiles in Rayleigh-B\'enard convection and logarithmic velocity profiles in Taylor-Couette flow are triggered by plume ejection \citep{ahl14,poe15,ost14bl}. These Refs. also support the notion that parts of the boundary layer can be turbulent, while others are not, and that the transition to the ultimate regime entails plume emission from the full extent of the boundary layer.

We will now further investigate the effect of the plumes on the logarithmic nature of the velocity profiles. As we have seen before, the roll structure that appears for counter-rotation of $a=0.2$ provides strong mean radial flows which cause plume emission away from either the inner or outer cylinder. 

In figure \ref{fig:kappa}(a) the angular velocity at the inner cylinder is shown. It can be seen that there is a significant variation in the shape of the profiles, and that profiles at the axial positions where a strong positive radial flow exists (top lines in figure \ref{fig:kappa}(c)) follow the shape of a logaritmic law more closely. This effect is even stronger at the outer cylinder in figure \ref{fig:kappa}(b), where profiles at an axial position with negative radial flow correspond best to a log-law. 

To quantify this phenomenom, the velocity profiles are fitted with the modified \citep{gro14} Prandtl-von K\'{a}rm\'{a}n law of the wall: $\omega^+=1/\kappa_{\omega} \ln{y^+} +B$, with both $\kappa_{\omega}$ and $B$ as fitting parameters. It should be noted that, as previously, a global torque is used to determine the normalisation of the profiles. The local torque can be different to the average torque, following the large-scale roll structures. This causes the imperfect matching with $\omega^+ = y^+$ and means that the value of the fitted parameter $\kappa_{\omega}$ can differ from the local value. The general shape of the profile is not affected, however. This shape, or specifically, the deviation from a log profile, is quantified by the standard error of the fitting parameter $\sigma(\kappa_{\omega})$. A low value of $\sigma(\kappa_{\omega})$ corresponds to a profile which closely follows a log-law. In figure \ref{fig:kappa}(d) it can be seen that for a mean wall normal velocity away from the cylinder, the value of $\kappa_{\omega}$ more closely approaches the known classical value of 0.40. Additionaly, the error of the fit is much smaller for positive values of the outward wall normal velocity as compared to negative values. We can summarise that for regions with strong radial flow away from a cylinder surface, plumes will emit from that cylinder, which in turn create logarithmic boundary layers. This confirms results from direct numerical simulations \citep{poe15,ost14bl}.
 
 \section{Conclusions}
 
In this work, we measured velocity profiles, the wind Reynolds number and characteristics of turbulent plumes in Taylor-Couette flow for a radius ratio of 0.5 and Taylor number of up to $6.2\cdot10^9$. The flow profiles show Taylor vortices for a Taylor number smaller than $\mathrm{Ta}\approx10^9$. The normalised profiles approach the Prandtl-von K\'{a}rm\'{a}n log-law, although at the highest  $\mathrm{Ta}=6.2\cdot10^9$ the log layer is not yet fully developed. They are in good agreement with DNS data from other work \citep{cho14,ost14pd,ost15}. Due to the strong curvature of this $\eta=0.5$ setup, a large difference between the azimuthal velocity $u^+$ and the angular velocity $\omega^+$ arises. The angular velocity $\omega^+$ resembles a log-law more closely, as suggested by \citet{gro14}, based on their Navier-Stokes based theoretical considerations.

Because of the late onset of the ultimate regime for $\eta=0.5$, the measurements with $\mathrm{Ta}$ up to $6.2\cdot10^9$ are in the classical turbulent regime. For the first time the wind Reynolds number has been measured in the classical regime of Taylor-Couette flow and it indeed follows the theoretically predicted classical scaling of $\mathrm{Re_w} \propto \mathrm{Ta}^{3/7}$.

Moreover, we focused on the interplay between rolls, turbulent plumes and logarithmic velocity profiles. At a strong driving of  $\mathrm{Ta}=4.2\cdot10^9$ no significant coherent structures exists for pure inner cylinder rotation, but roll structures appear for counter-rotation at $a=-\omega_o/\omega_i=0.2$. This behaviour has been observed previously for different values of $\eta$ and $\mathrm{Ta}$, and shows the important part that rolls play in momentum transfer between the two cylinders.

For inner cylinder rotation only, strong outward plumes are visible in the radial velocity. In the azimuthal velocity, the plumes mainly go inwards. The exact mechanism causing this difference is yet to be elucidated. For counter-rotation, the roll structures strongly influence the direction of the plumes. There is a direct correspondence between the direction of the mean radial flow and the direction of the plumes at a certain position. The plume velocity in the radial flow profiles increases with Taylor number, and has an approximately constant value of one quarter of the inner cylinder velocity for pure inner cylinder rotation.

Lastly, by quantifying the correspondence of the angular velocity profiles to a log-law for several axial positions, it was found that in regions with strong radial flow away from a cylinder surface, plumes will emit from that cylinder, which in turn create logarithmic boundary layers.

This work confirms predictions about velocity profiles and the scaling of the wind Reynolds number and sheds new light on the characteristics and the role of plumes in Taylor-Couette flow for a radius ratio $\eta=0.5$, and hopefully will spark further research into these intricate phenomena.

\section*{Acknowledgments}
We thank the technicians of the Department of Aerodynamics and Fluid Mechanics at the BTU for their technical support. This work was financially supported by the European High-Performance Infrastructures in Turbulence (EuHIT), an European Research Council (ERC) Advanced Grant, and the Simon Stevin Prize of the Technology Foundation STW of The Netherlands.

\bibliographystyle{jfm}

\end{document}